\documentclass[final,leqno]{siamltex}

\usepackage{graphicx} 

\usepackage{subfigure}
\usepackage{amsmath}
\usepackage{amsfonts}
\usepackage[ansinew]{inputenc}

\newtheorem{lem}{Lemma}
\newtheorem{thm}{Theorem}

\title{Kernels for linear time invariant system identification}

\author{Francesco Dinuzzo \thanks{Max Planck Institute for Intelligent Systems, Spemannstrasse 38, 72076 Tübingen, Germany ({\tt fdinuzzo@tuebingen.mpg.de}).}}

\begin{document}

\maketitle

\begin{abstract}
In this paper, we study the problem of identifying the impulse response of a linear time invariant (LTI) dynamical system from the knowledge of the input signal and a finite set of noisy output observations. We adopt an approach based on regularization in a Reproducing Kernel Hilbert Space (RKHS) that takes into account both continuous and discrete time systems. The focus of the paper is on designing  spaces that are well suited for temporal impulse response modeling. To this end, we construct and characterize general families of kernels that incorporate system properties such as stability, relative degree, absence of oscillatory behavior, smoothness, or delay. In addition, we discuss the possibility of automatically searching over these classes by means of kernel learning techniques, so as to capture different modes of the system to be identified.
\end{abstract}



\pagestyle{myheadings}
\thispagestyle{plain}
\markboth{Francesco Dinuzzo}{Kernels for LTI system identification}

\section{Introduction}

Identification of LTI systems is a fundamental problem in science and engineering  \cite{Ljung99}, which is most commonly solved by fitting families of parametric models via Maximum Likelihood and choosing the final one by means of statistical model selection techniques. While such an approach is well-established and has been proven to be successful in a variety of circumstances, several works have shown that even better performances can be obtained by balancing the data fitting with proper complexity control or regularization. For instance, a recent line of research, that adopts a state-space perspective, promotes the use of penalties based on the nuclear norm that encourage a low McMillan degree, see e.g. \cite{Liu09, Mohan10}. Another line, following  \cite{Pillonetto10}, adopts an approach based on Bayesian estimation of Gaussian Processes \cite{Rasmussen06} where a suitable prior is defined over possible impulse responses of the system. Equivalently, this amounts to penalize the data fitting with a kernel-based regularization of the impulse response \cite{Chen12}, which naturally poses the question of which kernels are best for the purpose of impulse response identification.


In this paper, we propose a systematic study of the problem of kernel selection for LTI system identification from a functional analytic perspective. We start by formulating LTI system identification as an inverse problem to be solved by means of regularization in a function space, in the spirit of \cite{Tikhonov77,Wahba90}. Such point of view provides a unified algorithmic framework that allows to take into account both discrete and continuous time systems. Moreover, it permits to go beyond the standard assumption of uniform time sampling, allowing for arbitrary sampling of the output signal. Knowledge about properties of the system can be incorporated naturally by exploiting standard characterizations of the impulse response.  Without loss of generality, we focus on SISO (single input single output) LTI system identification, where the goal is to reconstruct a scalar impulse response function from the knowledge of the input signal and a finite set of output measurements. Nevertheless, the ideas presented in this paper are general enough to be extendable to more complex and structured problems.



A central contribution of this paper is to characterize families of function spaces that encode those properties that are specific to impulse responses of dynamical systems, such as causality, stability, absence of oscillations, relative degree, or delay. In particular, we show how these structural properties can be enforced by designing suitable kernel functions. Importantly, we also provide theoretical support to the empirical evidence that standard translation invariant kernels such as the Gaussian RBF are not well-suited for modeling impulse responses of stable dynamical systems. Finally, we also discuss the possibility of optimizing the kernel by means of methodologies such as multiple kernel learning (MKL) \cite{Bach04, Micchelli05}, suggesting that these techniques can be rather appropriate in the context of system identification.

\section{Kernel based regularization for LTI system identification}

In order to handle both continuous and discrete-time system in a unified framework, we refer to an abstract \emph{time set} $\mathcal{T}$, that is a sub-group of $(\mathbb{R},+)$.

A dynamical system is called \emph{linear time invariant} (LTI) if, for any input signal $u:\mathcal{T}\rightarrow \mathbb{R}$, the output signal $y:\mathcal{T}\rightarrow \mathbb{R}$ is generated according to a convolution equation
\[
y(t) = (h \ast u)(t) = \int_{\mathcal{T}}u(\tau)h(t-\tau)d\tau,
\]
\noindent where $h:\mathcal{T}\rightarrow \mathbb{R}$ is the \emph{impulse response}. Depending on the nature of the time set, the symbol $\int_{\mathcal{T}}$ has to be interpreted as an integral, a series, or simply a sum.

In the following, we study the problem of identifying the impulse response, assuming availability of the input signal and a finite dataset of (noisy) output measurement pairs
\[
\mathcal{D} = \left\{(t_1, y_1), \ldots, (t_{\ell}, y_{\ell})\right\}.
\]
\noindent The problem can be tackled by means of regularization techniques, based on minimization problems of the form
\begin{equation}\label{EQUA01}
\min_{h\in\mathcal{H}}\left(\sum_{i=1}^{\ell} L\left(y_i,(h \ast u)(t_i)\right)+\frac{\lambda}{2}\|h\|_{\mathcal{H}}^2\right),
\end{equation}
\noindent where $\mathcal{H}$ is a Hilbert space of functions, $L$ is a loss function that is convex and continuous w.r.t. the second argument, and $\lambda > 0$ is a regularization parameter. We assume that the input signal and the space $\mathcal{H}$ are such that all the point-wise evaluated convolutions are bounded linear functionals, namely, for all $i =1, \ldots, \ell$, there exists a finite constant $C_i$ such that
\[
|(h \ast u)(t_i)| \leq C_i  \|h\|_{\mathcal{H}}, \quad \forall h \in \mathcal{H}.
\]
\noindent Then, there exist unique representers $w_i$ such that
\[
(h \ast u) (t_i) = \langle h, w_i \rangle_{\mathcal{H}}.
\]
\noindent In addition, one can show that any optimal solution of (\ref{EQUA01}) can be expressed in the form
\begin{equation}\label{EQUA02}
h^* = \sum_{i=1}^{\ell} c_i w_i.
\end{equation}
\noindent This result, known as the \emph{representer theorem}, see e.g. \cite{Kimeldorf71, Dinuzzo12}, shows that the regularization problem (\ref{EQUA01}) reduces to determining a vector of coefficients $c_i$ of the same dimension of the number of observations. More precisely, an optimal vector of coefficients $c \in\mathbb{R}^{\ell}$ can be obtained by solving the following convex optimization problem
\begin{equation}\label{EQUA03}
\min_{c \in \mathbb{R}^{\ell}} \left(\sum_{i=1}^{\ell} L\left(y_i,(\mathbf{K}c)_i\right)+\frac{\lambda}{2}c^T\mathbf{K}c\right)
\end{equation}
\noindent where the entries of the \emph{kernel matrix} $\mathbf{K}$ are given by
\[
\mathbf{K}_{ij} = \langle w_i, w_j \rangle_{\mathcal{H}}.
\]

\subsection{Reproducing Kernel Hilbert Spaces}

Reproducing Kernel Hilbert Spaces \cite{Aronszajn50} are a family of Hilbert spaces that enjoy particularly favorable properties from the point of view of regularization. The concept of RKHS is strongly linked with that of \emph{positive semidefinite kernel}. Given a non-empty set $\mathcal{X}$, a positive semidefinite kernel is a symmetric function $K:\mathcal{X} \times \mathcal{X} \rightarrow \mathbb{R}$ such that
\[
\sum_{i=1}^{\ell}\sum_{j=1}^{\ell}c_i c_j K(x_i,x_j) \geq 0, \qquad \forall (x_i,c_i) \in \left(\mathcal{X},\mathbb{R}\right).
\]

A RKHS is a space of functions $h:\mathcal{X}\rightarrow \mathbb{R}$ such that point-wise evaluation functionals are bounded. This means that, for any $x \in \mathcal{X}$, there exists a finite constant $C_x$ such that
\[
|h(x)| \leq C_x \|h\|_{\mathcal{H}}, \quad \forall h \in \mathcal{H}.
\]
\noindent Given a RKHS, it can be shown that there exists a unique symmetric and positive semidefinite kernel function $K$ (called the \emph{reproducing kernel}), such that the so-called \emph{reproducing property} holds:
\[
h(x) = \langle h, K_x \rangle_{\mathcal{H}}, \qquad \forall \left(x,h\right) \in \mathcal{X} \times \mathcal{H},
\]
\noindent where the \emph{kernel sections} $K_x$ are defined as
\[
K_x(y) = K(x,y), \qquad \forall y \in \mathcal{X}.
\]
\noindent The reproducing property states that the representers of point-wise evaluation functionals coincide with the kernel sections. Starting from the reproducing property, it is also easy to show that the representer of any bounded linear functional $L$ is a function $K_L \in \mathcal{H}$ such that
\[
K_L(x) = L K_x, \qquad  \forall x \in \mathcal{X}.
\]
\noindent Therefore, in a RKHS, the representer of any bounded linear functional can be obtained explicitly in terms of the reproducing kernel.

With reference to the problem (\ref{EQUA01}), we are interested in estimating functions defined over the time set $\mathcal{X} = \mathcal{T}$. By expressing the representers in terms of the kernel, the optimal solution (\ref{EQUA02}) can be rewritten as
\[
h^*(t) = \sum_{i=1}^{\ell} c_i (u \ast K_t)(t_i).
\]

\noindent The entries of the kernel matrix $\mathbf{K}$ that appears in problem (\ref{EQUA03}) can be computed as
\[
\mathbf{K}_{ij} = \int_{\mathcal{T}}\int_{\mathcal{T}} u(t_i-\tau_1) u(t_j-\tau_2) K(\tau_1,\tau_2) d\tau_1 d\tau_2.
\]

For discrete-time problems, the integral reduces to sums, whereas for continuous-time problems they can be approximated by means of numerical integration techniques. For a variety of kernel functions, the continuous-time integrals can be even computed in closed form, provided that the input signal is known to have a sufficiently simple expression.

\section{Enforcing basic system properties}

By searching over a RKHS, the impulse response to be synthesized is automatically constrained to be point-wise well-defined and bounded over compact time sets. In this section, we show how several other important properties of the impulse response can be enforced by adopting suitable kernel functions.

\subsection{Causality}

A dynamical system is said to be causal if the value of the output signal at a certain time instant $T$ does not depend on values of the input in the future (for $t > T$). Knowledge about causality is virtually always incorporated in the model of a dynamical system. This is done already when the signals are classified as inputs or outputs of the system: the value of the output signals at a certain time is not allowed to depend on the values of the input signals in the future.

For a LTI system, causality is equivalent to vanishing of the impulse response for negative times, namely
\begin{equation}\label{EQUA04}
h(t) = 0, \quad \forall t < 0,\quad \forall h \in \mathcal{H}.
\end{equation}
The following Lemma characterizes those RKHS that contain causal impulse responses, with a simple condition on the kernel function.

\begin{lem}\label{LEM01}
The RKHS $\mathcal{H}$ contains only causal impulse responses if and only if the reproducing kernel satisfy
\begin{equation}\label{EQUA05}
K(t_1,t_2) = H(t_1)H(t_2)\widetilde{K}(t_1,t_2).
\end{equation}
\noindent where $H(t)$ is the Heaviside step function defined as
\[
H(t) = \left\{
\begin{array}{ll}
1, & t \geq 0 \\
0, & \textrm{else}
\end{array}
\right.
\]
and $\widetilde{K}$ is a kernel defined for non-negative times.
\end{lem}

The simple statement of Lemma \ref{LEM01} already shows that the kernels needed for modeling impulse responses of dynamical systems are quite different from the typical kernels used for curve fitting. In order to encode a “privileged” direction in the time flow, they have to be asymmetric on the real line, and can also be discontinuous.

\subsection{Stability}

System stability is an important information that is often known to be satisfied by the system under study and should be always incorporated whenever available. Perhaps, the most intuitive notion of stability is the so called BIBO (Bounded Input Bounded Output) condition that can be expressed as
\[
\| u \|_{\infty} < +\infty \Rightarrow \| y \|_{\infty} < +\infty,
\]

\noindent where $\| \cdot \|_{\infty}$ denotes the $L^{\infty}$ norm. BIBO stability entails that the output signal cannot diverge when the system is excited with a bounded input signal. While ensuring stability for methods based on state-space models requires special techniques, see e.g \cite{VanGestel01, Siddiqi08}, the RKHS regularization framework can handle this constraint very easily. Indeed, it is well-known that for a LTI system, BIBO stability is equivalent to integrability of the impulse response:
\[
 \int_{ \mathcal{T}}\left| h(t) \right| dt< +\infty.
\]

\noindent Hence, in order to encode stability, it is sufficient to characterize those RKHS that contain only integrable impulse responses. The following Lemma gives a necessary and sufficient condition (see e.g. \cite{Carmeli06}):
\begin{lem}\label{LEM02}
The RKHS $\mathcal{H}$ is a subspace of $L^1( \mathcal{T})$ if and only if
\[
\int_{ \mathcal{T}}\left|\int_{\mathcal{T}}K(t_1,t_2) h(t_1)dt_1 \right| dt_2 < + \infty, \quad \forall h \in L^{\infty}( \mathcal{T}).
\]
\end{lem}

\noindent We can talk about \emph{stability of the kernel}, with reference to kernels that satisfy the conditions of Lemma \ref{LEM02}. It can be easily verified that integrability of the kernel is a sufficient condition for $\mathcal{H}$ to be a subspace of $L^1( \mathcal{T})$.

\begin{lem}\label{LEM03}
If $K \in L^1(\mathcal{T}^2)$, then $\mathcal{H}$ is a subspace of $L^1( \mathcal{T})$.
\end{lem}

\noindent It is worth observing that the condition of Lemma \ref{LEM03} is also necessary for nonnegative-valued kernels (i.e. such that $K(t_1,t_2) \geq 0$, for all $t_1,t_2$), as it can be seen by simply setting $h(t) =1$ in Lemma \ref{LEM02}.

\subsection{Delay}

Let $D$ denote the inferior of the time instants where the impulse response is not equal to zero:
\[
D := \inf\left\{t \in  \mathcal{T}: h(t) \neq 0 \right\}.
\]

\noindent By causality, $D$ has to be nonnegative. If it is strictly positive, then the system is said to exhibit an input-output \emph{delay} equal to $D$, meaning that $y(\tau)$ does not depend on $u(t)$ for any $t > \tau- D$. The knowledge of the delay $D$ can be easily incorporated in the kernel function.

\begin{lem}\label{LEM04}
Every impulse response $h \in \mathcal{H}$ have a delay equal to $D$ if and only if the reproducing kernel is in the form
\[
K_D(t_1,t_2) = K(t_1-D,t_2-D),
\]
\noindent with $K$ in the form (\ref{EQUA05}).
\end{lem}

\noindent If the value of $D$ is unknown in advance, it can be treated as a kernel design parameter to be estimated from the data.

\section{Kernels for continuous-time systems }

In this section, we focus on properties of continuous-time systems ($\mathcal{T} = \mathbb{R}$), such as smoothness of the impulse response and relative degree, and discuss how to enforce them by choosing suitable kernels.

\subsection{Smoothness}

Impulse responses of continuous-time dynamical systems are typically assumed to have some degree of smoothness. Without loss of generality, we focus on systems without delay (the delayed case can be simply handled via the change of variable discussed in Lemma \ref{LEM04}). Typically, we would like to have continuity of $h$ and a certain number of time derivatives, everywhere with the possible exception of the origin. Regularity of the impulse response at $t = 0$ is related to the concept of \emph{relative degree}, which is important enough to deserve an independent treatment (see the next subsection). Impulse responses with a high number of continuous derivatives corresponds to low-pass dynamical systems that attenuates high frequencies of the input signal. It is known that regularity of the kernel propagates to every function in the RKHS. Therefore, knowledge about smoothness of the impulse response can be directly expressed in terms of the kernel function, see e.g. \cite{Steinwart08}.

\begin{lem}\label{LEM05}
Let $\mathcal{H}$ denote a RKHS associated with the kernel in the form (\ref{EQUA05}) with $\mathcal{T} =\mathbb{R}$. If $\widetilde{K}$ is $k$-times continuously differentiable on $(0,+\infty)^{2}$, then the restriction of every function $h \in \mathcal{H}$ to $(0, +\infty)$ is $k$-times continuously differentiable. In addition, point-wise evaluated derivatives are continuous linear functionals, i.e. for all $t > 0$ and $i \leq k$, there exists $C < +\infty$ such that
\[
|h^{(i)}(t)| \leq C \|h\|_{\mathcal{H}}, \quad \forall h \in \mathcal{H}.
\]
\end{lem}

\subsection{Relative degree}

The relative degree of an LTI system is a quantity related to the regularity of the impulse response at $t =0$ (or $t = D$ in the delayed case). By causality, all the left derivatives of the impulse response (with the convention $h^{(0)} = h$) have to vanish:
\[
h^{(k)}(0^-) = 0, \quad \forall k \geq 0.
\]
\noindent On the other hand, the right derivatives may well be different from zero. Assuming existence of all the necessary derivatives, the relative degree of a LTI system is a natural number $k$ such that
\[
h^{(i)}(0^+) = 0, \quad \forall i < k, \quad h^{(k)}(0^+) \neq 0.
\]
If $h^{(i)}(0^+) = 0$ for all $i$, the relative degree is \emph{undefined}.

If the relative degree is $k$, then the $k$-th derivative of the impulse response at $t = 0$ is discontinuous. Let's represent the impulse response in the form $h(t) = H(t) h_+(t)$, where $H$ is the Heaviside step function, and assume that $h_+(t)$ admits at least $k$ right derivatives at $t = 0$. By using distributional derivatives and properties of the convolution, we have
\begin{align*}
& y^{(k+1)}(t)     = (h^{(k+1)} \ast u) (t)\\
                & = h^{(k)}(0^+) \left(\delta_0\ast u\right)(t)+ \int_{-\infty}^{t}u(\tau)h_+^{(k+1)}(t-\tau)d\tau \\
                & = h^{(k)}(0^+) u(t) + \int_{-\infty}^{t}u(\tau)h_+^{(k+1)}(t-\tau)d\tau.
\end{align*}

\noindent The $(k+1)$-th time derivative of the output is the first derivative that is \emph{directly} influenced by the input $u(t)$. Therefore, the system exhibits an input-output integral effect equivalent to a chain of $k$ integrators on the input of a system with relative degree one.

In many cases, knowledge about the relative degree is available thanks to simple physical considerations. Such knowledge can be enforced by designing the kernel according with the following Lemma.

\begin{lem}\label{LEM06}
Under the assumptions of Lemma \ref{LEM05}, every impulse response $h \in \mathcal{H}$ has relative degree greater or equal than $k$ if and only if
\begin{equation}\label{EQUA06}
\forall t \in \mathbb{R}, \quad \lim_{\tau\rightarrow 0^{+}}\widetilde{K}_t^{(i)}(\tau) = 0, \quad \forall i < k.
\end{equation}
\end{lem}

Hence, when the impulse response is searched within an RKHS, the relative degree of the identified system is directly related to the simple property (\ref{EQUA06}) of the kernel function. We can therefore introduce the concept of \emph{relative degree of the kernel}.

\subsection{Examples}

The simplest possible kernel of the form (\ref{EQUA06}) is the \emph{Heaviside kernel}
\[
K(t_1,t_2) = H(t_1)H(t_2),
\]
\noindent whose associated RKHS contains only step functions. This kernel has relative degree equal to one and is clearly not stable. As a second example, consider the \emph{exponential kernel}
\begin{equation}\label{EQUA07}
K(t_1,t_2) = H(t_1)H(t_2)e^{-\omega(t_1+t_2)}.
\end{equation}
\noindent This kernel is stable for any $\omega > 0$, infinitely differentiable everywhere, except over the lines $t_1 = 0$ and $t_2 = 0$, where it is discontinuous. Since $K$ is discontinuous, the relative degree is equal to one. The associated Hilbert space $\mathcal{H}$ contains exponentially decreasing functions. A third example is the TC (Tuned-Correlated) kernel \cite{Pillonetto10, Chen12} defined as
\begin{equation}\label{EQUA08}
K(t_1,t_2) = H(t_1)H(t_2)e^{-\omega\max\{t_1,t_2\}},
\end{equation}
\noindent which has relative degree equal to one and can be shown to be stable (see next section).

\section{Kernels for stable systems}

The exponential kernel defined in (\ref{EQUA07}) satisfies the sufficient condition of Lemma \ref{LEM03}, therefore the associated RKHS contains stable impulse responses of relative degree one (in fact, the space contains only stable exponential functions). Now, assume that a kernel $K_1$ with relative degree one is available. Then, we can easily generate a family of kernels of arbitrary relative degree via the following recursive procedure:
\[
K_{i+1}(t_1,t_2) = \int_{-\infty}^{t_1}\int_{-\infty}^{t_2}K_i(\tau_1,\tau_2)d\tau_1 d \tau_2, \quad i \geq 1.
\]
\noindent Unfortunately, such procedure does not preserve stability. Consider for example the exponential kernel (\ref{EQUA07}). Although $K_1$ is stable, all the other kernels $K_i$ with $i \geq 2$ do not satisfy the necessary condition of Lemma \ref{LEM02} (to see this, it is sufficient to choose $h = 1$) and are therefore not BIBO stable. In the following, we describe some alternative ways of constructing families of stable kernels.

\subsection{Mixtures of exponentially-warped kernels}

In this subsection, we discuss a technique to construct stable kernels of any relative degree. In order to obtain stability, we adopt a change of coordinates that maps $\mathbb{R}^+$ into the finite interval $[0, 1]$, and then use a kernel over the unit square $[0, 1]^2$. Let $G:[0, 1]^2\rightarrow \mathbb{R}$ denote a positive semidefinite kernel, and $h_{\omega }:\mathbb{R}^+ \rightarrow [0, 1]$ denote the exponential coordinate transformation
\[
h_{\omega }(t) = e^{-\omega t}.
\]
\noindent Then, we can construct a class of kernels defined as in (\ref{EQUA05}), where
\begin{equation}\label{EQUA09}
\widetilde{K}(t_1,t_2) = (t_1t_2)^k\int_{\mathbb{R}^+}G(h_{\omega}(t_1),h_{\omega}(t_2))d\mu(\omega), \quad k \in \mathbb{N},
\end{equation}
\noindent and $\mu$ is a probability measure. If $G(h_{\omega }(t_1),h_{\omega}(t_2))$ is a kernel with relative degree one, we can check, using Lemma \ref{LEM06}, that the kernel (\ref{EQUA09}) has relative degree $k+1$. To ensure BIBO stability, the mass of $\mu(\omega)$ should not be concentrated around zero and the kernel $G$ must vanish sufficiently fast around the origin. The following Lemma gives a sufficient condition.

\begin{lem}\label{LEM07}
Let $G:[0, 1]^2\rightarrow \mathbb{R}$ denote a kernel such that
\begin{equation}\label{EQUA10}
|G(s_1,s_2)| \leq C s_1s_2, \quad \forall (s_1,s_2) \in [0, 1]^2.
\end{equation}
\noindent If the support of $\mu$ does not contain the origin, then the kernel (\ref{EQUA09}) is BIBO stable for all $k \in \mathbb{N}$.
\end{lem}

A particular case of this construction is the TC kernel (\ref{EQUA08}), where $G(s_1,s_2) = \min\{s_1, s_2\}$, and (\ref{EQUA10}) is satisfied with $C =1$. Another example is the cubic \emph{stable spline} kernel \cite{Pillonetto10}, obtained by choosing $G$ as the cubic spline kernel (that can be also seen as the covariance function of an integrated Wiener process on $\mathbb{R}^+$):
\[
G(s_1,s_2) = \frac{s_1s_2 \min\{s_1,s_2\}}{2}-\frac{\min\{s_1,s_2\}^3}{6}.
\]
\noindent A simple calculation shows that condition (\ref{EQUA10}) is satisfied with $C = 1/3$. By using (\ref{EQUA09}), we can generate a class of stable kernels of arbitrary relative degree. For example, the kernel
\[
\widetilde{K}(t_1,t_2) = (t_1t_2)^{k} e^{-\omega\max\{t_1,t_2\}}
\]
\noindent is obtained by choosing $G(s_1,s_2) = \min\{s_1, s_2\}$ and $\mu$ as the unit mass on a certain frequency $\omega > 0$. This kernel is stable and has relative degree equal to $k+1$.

\subsection{Kernels for relaxation systems}

Many real-world systems, such as reciprocal electrical networks whose energy storage elements are of the same type, or mechanical systems in which inertial effects may be neglected, have the property that the impulse response never exhibits oscillations. \emph{Relaxation systems}, see e.g. \cite{Willems72}, are dynamical systems whose impulse response is a so-called \emph{completely monotone} function. An infinitely differentiable function $f:\mathbb{R}^+ \rightarrow \mathbb{R}$ is called completely monotone if
\[
(-1)^{n} f^{(n)}(t) \geq 0, \quad \forall n \in \mathbb{N}, \quad t > 0.
\]

The following characterization of completely monotone functions \cite{Bernstein28,Widder41} allows to generalize the basic exponential kernel defined in (\ref{EQUA07}).

\begin{thm}[Bernstein-Widder]\label{THM01}
An infinitely differentiable  real-valued function $f$ defined on the real line is completely monotone if and only if there exists a non-negative finite Borel measure $\mu$ on $\mathbb{R}^+$ such that
\[
f(t) = \int_{\mathbb{R}^+}e^{-t\omega}d\mu(\omega).
\]
\end{thm}

\noindent In view of this last theorem, completely monotone functions are characterized as mixture of decreasing exponentials or, in other words, as Laplace transforms of non-negative measures. Let $f$ denote a completely monotone function, and consider the family of functions of the form
\begin{equation}\label{EQUA11}
K(t_1,t_2) = H(t_1)H(t_2) f(t_1+t_2).
\end{equation}

\noindent By Theorem \ref{THM01}, we can easily verify that (\ref{EQUA11}) defines a positive semidefinite kernel:
\begin{align*}
&\sum_{i=1}^{\ell}\sum_{j=1}^{\ell}c_i c_jK(t_i,t_j) =\\
& = \sum_{i=1}^{\ell}\sum_{j=1}^{\ell}c_i c_j H(t_i)H(t_j)\int_{\mathbb{R}^+}e^{-t_i \omega}e^{-t_j\omega}d\mu(\omega)\\
& = \int_{\mathbb{R}^+}\left(\sum_{i=1}^{\ell}c_i H(t_i)e^{-t_i \omega}\right)^2d\mu(\omega) \geq 0.
\end{align*}

\noindent Clearly, not every function in the associated RKHS is a completely monotone impulse response. However, all the kernel sections $K_t$ are completely monotone. Now, observe that, unless $f = 0$, the relative degree of kernel (\ref{EQUA11}) is always one. Indeed, if the relative degree is greater than one, then we have $ f(t) = K_t(0^+) = 0$, for all $t \in \mathbb{R}$. By using Lemma \ref{LEM03}, we can check that, when the support of $\mu$ does not contain the origin, the kernel (\ref{EQUA11}) is BIBO stable:
\begin{align*}
& \int_{\mathbb{R}^2}\left|K(t_1,t_2)\right| dt_1 dt_2 =\\
& = \int_{\epsilon}^{+\infty}\left(\int_{\mathbb{R}^+}e^{-t_1 \omega}dt_1\right)\left(\int_{\mathbb{R}^+}e^{-t_2 \omega}dt_2\right) d\mu(\omega)\\
& = \int_{\epsilon}^{+\infty} \frac{d\mu(\omega)}{\omega^2} \leq \frac{1}{\epsilon^2}\int_{\epsilon}^{+\infty}d\mu(\omega) < + \infty.
\end{align*}

\noindent On the other hand, if the support of $\mu$ contains the origin, we may obtain unstable kernels. For instance, when $\mu$ is the unitary mass centered at the origin, we obtain the Heaviside kernel $H(t_1)H(t_2)$, which is not stable.

Finally, observe that not all the kernels of the form (\ref{EQUA11}) that vanishes when $t_1$ or $t_2$ tend to $+\infty$ are stable, as shown by the following counterexample:
\[
K(t_1,t_2) = \frac{H(t_1)H(t_2)}{1+(t_1+t_2)^2}.
\]

\noindent This kernel is indeed of the type (\ref{EQUA11}), since the function $(1+t^2)^{-1}$ is completely monotone. However, the necessary condition of Lemma \ref{LEM02} is not satisfied with $h =1$:
\begin{align*}
& \int_{\mathbb{R}^+}\left|\int_{\mathbb{R}^+}\frac{dt_1}{1+(t_1+t_2)^2}\right|dt_2 \\
& = \int_{\mathbb{R}^+}\left(\frac{\pi}{2}-\tan^{-1}\left(t_2\right)\right)dt_2\\
& =  1 + \frac{1}{2}\log(1+t_2^2)\bigg|_{0}^{+\infty} = +\infty.
\end{align*}

\subsection{Translation invariant kernels are not stable}

In contrast with (\ref{EQUA11}), consider now kernels the form
\begin{equation}\label{EQUA12}
K(t_1,t_2) = H(t_1)H(t_2) f(t_1-t_2).
\end{equation}

The following classical result \cite{Schoenberg38} characterizes the class of functions such that $f(t_1-t_2)$ is a positive semidefinite kernel.

\begin{thm}[Schoenberg]\label{THM2}
Let $f: \mathbb{R} \rightarrow \mathbb{R}$ denote a continuous function. Then, $f(t_1-t_2)$ is a positive semidefinite kernel if and only if there exists a non-negative finite Borel measure $\mu$  on $\mathbb{R}^+$ such that
\[
f(t) = \int_{\mathbb{R}^+}\cos(t\omega)d\mu(\omega).
\]
\end{thm}

\noindent Hence, when $f$ is the cosine transform of a non-negative measure, the functions of the form (\ref{EQUA12}) are positive semidefinite kernels, since they are the product of the Heaviside kernel and a positive semidefinite kernel.

The family includes oscillating functions
\[
f(t_1-t_2) = \sum_{i=1}^{m}d_i \cos(\omega_i(t_1-t_2)),
\]
\noindent but also widely used kernels like the Gaussian
\[
f(t_1-t_2) = e^{-\omega(t_1-t_2)^2}.
\]
\noindent In view of their popularity in regression, one might be tempted to adopt these kernels for system identification. However, it turns out that, unless $f = 0$, kernels defined by (\ref{EQUA12}) are never stable.

\begin{lem}\label{LEM08}
The only BIBO stable kernel of the form (\ref{EQUA12}) is $K = 0$.
\end{lem}

In view of Lemma \ref{LEM08}, we can conclude that the class of kernels (\ref{EQUA12}) is not well-suited for identification of stable systems.

\section{Optimizing the kernel}

In the previous section, we have encountered families of kernels parameterized by a non-negative measure $\mu$, such as (\ref{EQUA09}) and (\ref{EQUA11}). It is therefore natural to consider the idea of searching over these classes by optimizing the measure $\mu$ simultaneously with the impulse response $h$. Since the measure $\mu$ defines the kernel, and the kernel in turn identifies the RKHS, such simultaneous optimization amounts to searching for $h$ into the union of an infinite family of spaces $\mathcal{H}_{\mu}$.

A possible approach to address such joint optimization has been studied in \cite{Argyriou05}, based on the solution of problems of the form
\begin{equation}\label{EQUA13}
\min_{\mu \in \mathcal{P}(\Omega)} \left[\min_{h \in \mathcal{H}_{\mu}} \left(\sum_{i=1}^{\ell}L(y_i, (h \ast u)(t_i))+\frac{\lambda}{2} \|h\|_{\mathcal{H}_{\mu}}^2\right)\right],
\end{equation}
\noindent where $\mathcal{P}(\Omega)$ is the class of probability measures over a compact set $\Omega$. Here, the fact that we search over probability measures (instead of generic non-negative measures) is necessary to make the problem well-posed.

Remarkably, one can still characterize an optimal solution of problem (\ref{EQUA13}) by means of a finite dimensional parametrization. First of all, for any fixed $\mu$, the standard representer theorem applies to the inner minimization problem, so that the optimal $h^*$ can be expressed in the form (\ref{EQUA02}). In addition, it turns out that there exists an optimal discrete measure $\mu^*$ with mass concentrated at no more than $\ell +1$ points:
\[
\mu^* = \sum_{i=1}^{\ell+1} d_i \delta_{\omega_i},
\]
\noindent and therefore, there exists also an optimal kernel that can be written as a finite convex combination of basis kernels:
\[
K^* = \sum_{i=1}^{\ell+1} d_i K_{\omega_i}.
\]
\noindent For instance, by searching over the class (\ref{EQUA11}) (where the support of $\mu$ is restricted to a compact set of the form $\left[0, \omega_{M}\right]$), one obtains an optimal kernel of the type
\[
K^*(t_1,t_2) = H(t_1)H(t_2)\sum_{i=1}^{\ell+1}d_i e^{-\omega_i(t_1+t_2)},
\]
\noindent namely a convex combination of decreasing exponential kernels with at most $\ell+1$ different rates. Unfortunately, the optimal decay rates $\omega_i$ solve a non-convex optimization problem where non-global local minimizers are possible.


A relaxation of problem (\ref{EQUA13}) consists in fixing a set of $m$ parameters $\omega_i$ over a sufficiently fine grid, and directly searching the measure $\mu$ into the finite-dimensional family
\[
\mathcal{P}_m(\Omega) =\left\{ \mu \in \mathcal{P}(\Omega): \mu = \sum_{i=1}^{m} d_i \delta_{\omega_i}\right\}.
\]

The relaxed problem
\begin{equation}\label{EQUA14}
\min_{\mu \in \mathcal{P}_m(\Omega)} \left[\min_{h \in \mathcal{H}_{\mu}} \left(\sum_{i=1}^{\ell}L(y_i, (h \ast u)(t_i))+\frac{\lambda}{2} \|h\|_{\mathcal{H}_{\mu}}^2\right)\right],
\end{equation}
\noindent boils down to a multiple kernel learning (MKL) problem, after application of the representer theorem:
\begin{align*}
& \min_{d \in \Delta_m}\left[\min_{c \in \mathbb{R}^{\ell}} \sum_{i=1}^{\ell} L\left(y_i,\left(\mathbf{K}c\right)_i\right)+\frac{\lambda}{2}c^T\mathbf{K}c\right],\\
& \textrm{subject to} \quad \mathbf{K} = \sum_{k=1}d_k\mathbf{K}_k,
\end{align*}
\noindent where $\Delta_m$ is the standard simplex in $\mathbb{R}^m$, and $\mathbf{K}_k$ are the kernel matrices associated with the different basis kernels $K_{\omega_k}$. Albeit not jointly convex, (\ref{EQUA14}) can be solved globally for many loss functions.


%

\subsection{Illustrative Example}

The following experiment shows that solving (\ref{EQUA14}) is an effective way to perform continuous LTI system identification with generic time sampling. In addition, it illustrates the advantages of incorporating information such as the relative degree of the system in the kernel design.

\begin{figure}
\centerline{ \includegraphics[width=1\columnwidth]{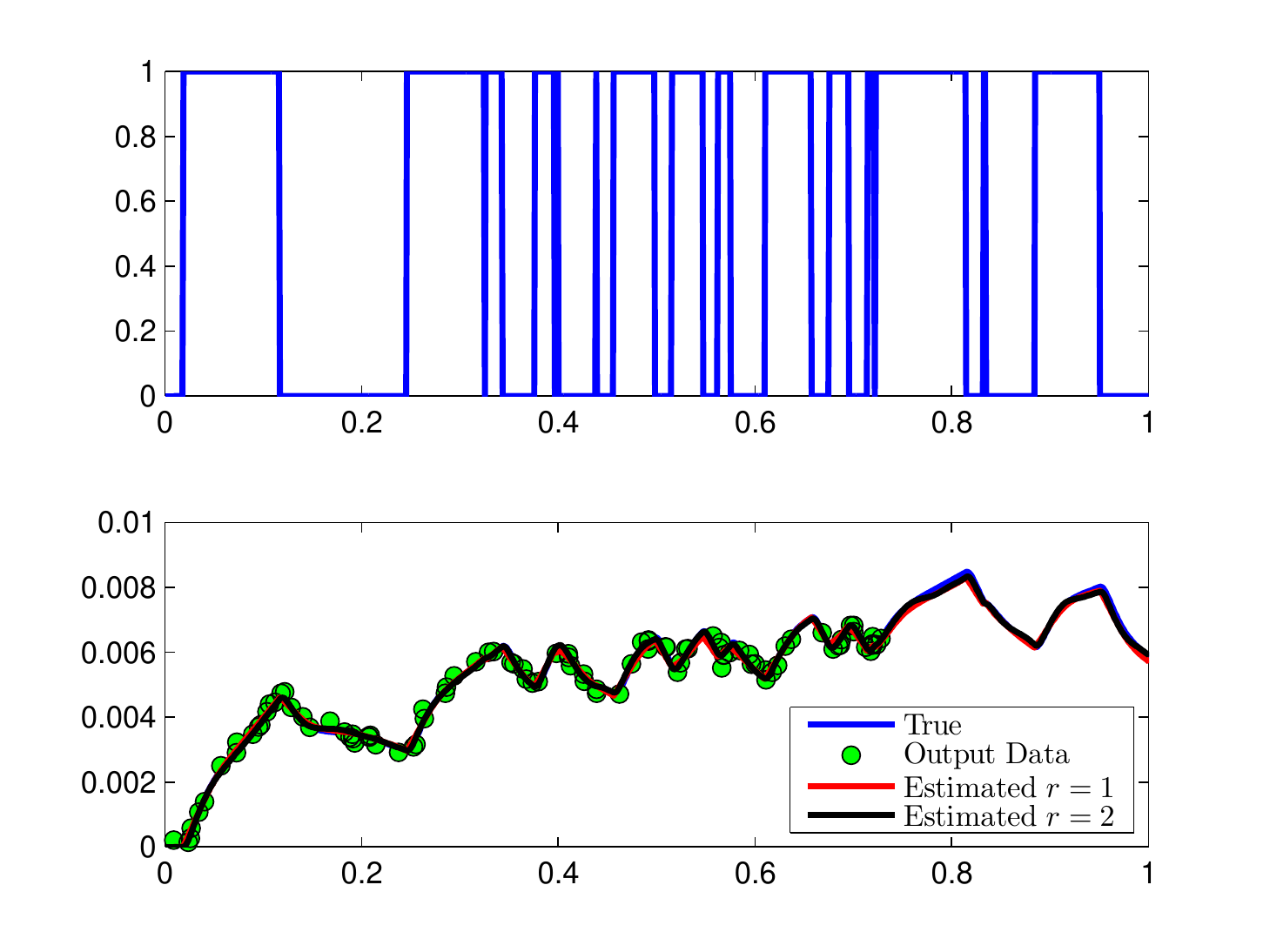}}
\caption{Input and output signals (blue lines), output measurements (green circles), and output estimates.}\label{FIG01}
\end{figure}

Consider the following bimodal impulse response with relative degree $r =2$:
\[
h(t) = H(t) \left(e^{-\omega_1 t} +  A e^{-\omega_2 t}\right) t,
\]
\noindent where $\omega_1 = 10$, $\omega_2 = 100$, and $A = 20$. We generate a binary input signal where the switching instants $t_i^U$ are randomly uniformly drawn from the interval $[0, 1]$ (top panel of Figure \ref{FIG01}). A set of $\ell = 100$ time instants $t_i$ are then drawn uniformly from the interval $[0, 0.75]$, and a vector of noisy output measurements is generated as
\[
y = y^0 + \epsilon,
\]
\noindent where $y_i^0 = (u \ast h)(t_i)$, and the $\epsilon_i$ are independent zero-mean Gaussian random variables with standard deviation $\sigma = 0.1 \cdot \textrm{std}(y^0)$.

\begin{figure}
\centerline{ \includegraphics[width=1\columnwidth]{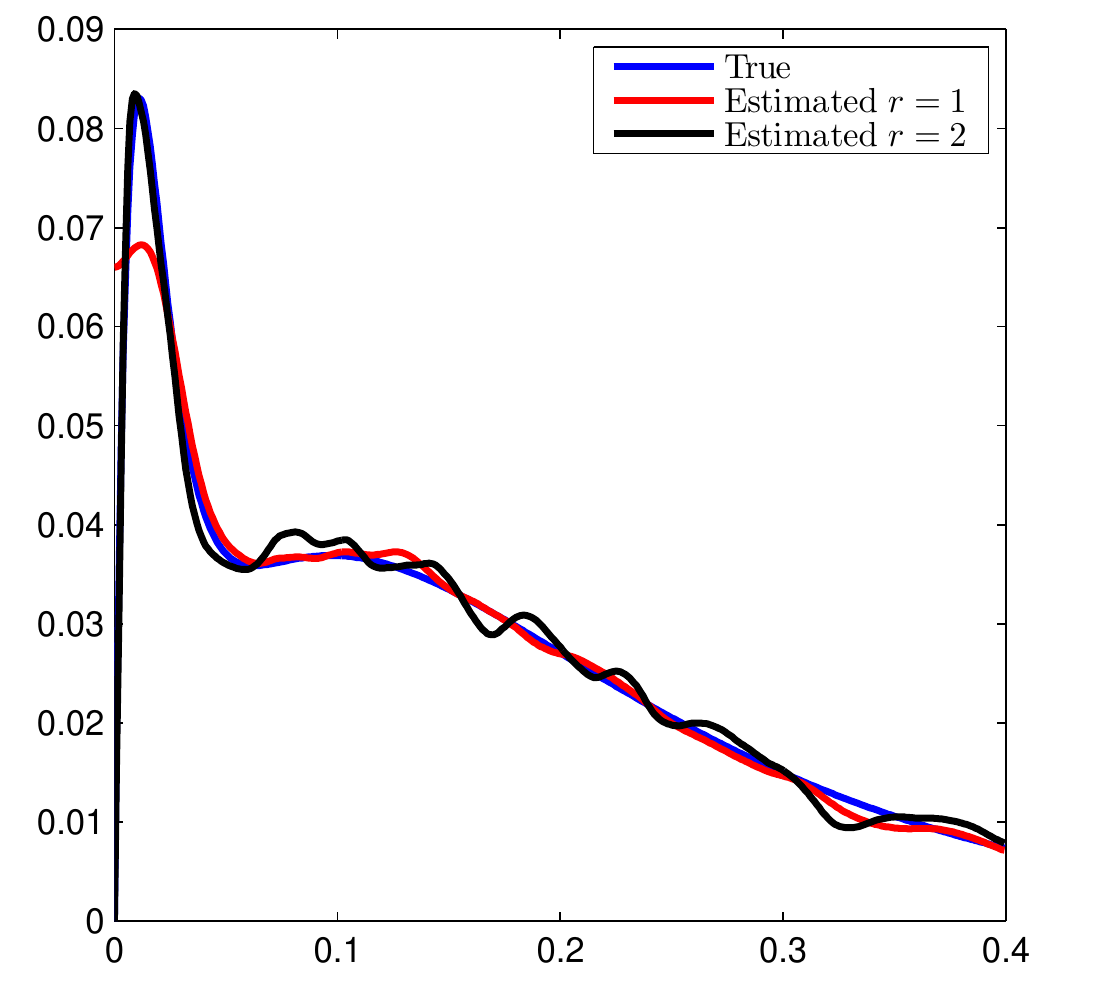}}
\caption{True and estimated impulse responses.}\label{FIG02}
\end{figure}

We run 50 independent experiments with different realizations of the output noises $\epsilon_i$. For each experiment, we solve the MKL problem (\ref{EQUA14}) with a least square loss function using the RLS2 tool described in \cite{Dinuzzo10}, and $m = 40$ basis kernels of the form
\[
K_{\omega_k}(t_1,t_2) = H(t_1)H(t_2) (t_1 t_2)^{r-1} e^{-\omega_k\max\{t_1,t_2\}}.
\]
\noindent The $\omega_k$ are chosen on a logarithmic scale in the interval $[10^0, 10^3]$, and $r$ is fixed to either 1 or 2. The regularization parameter $\lambda$ is selected automatically by minimizing a Generalized Cross Validation (GCV) score \cite{Craven79} over a logarithmic grid.  Figure \ref{FIG01} shows the input and the output signals, together with a set of output measurements and the estimated output signals for one of the 50 experiments. Both $r=1$ and $r=2$ yield excellent estimates of the output signal, also in the region where no measurements are available ($t > 0.75$). The true and estimated impulse responses are plotted in Figure \ref{FIG02}, showing that the kernels with $r =2$ are able to capture much better the fast mode. By inspecting the coefficients $d_i$, one can observe that indeed only two of them are different from zero, capturing the two dominant frequencies of the system.

\begin{figure}
\centerline{ \includegraphics[width=1\columnwidth]{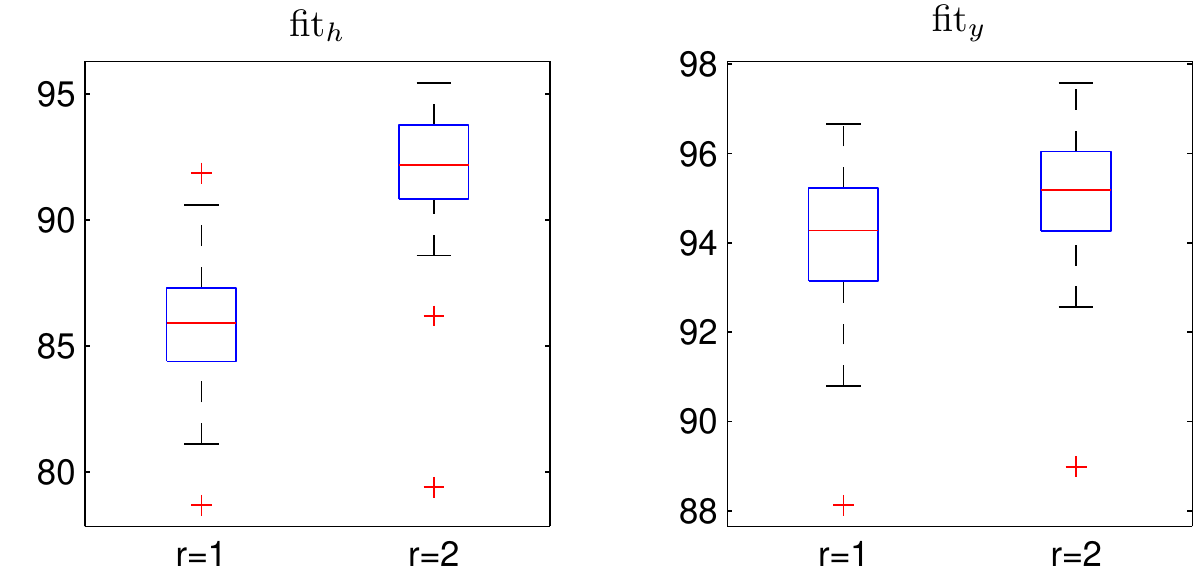}}
\caption{Boxplots of $\textrm{fit}_h$ (left) and $\textrm{fit}_y$ (right) for the 50 experiments.}\label{FIG03}
\end{figure}

For each of the 50 experiments, performances are evaluated according to the following scores:
\[
\textrm{fit}_h = 100\left(1-\sqrt{\frac{\int_{0}^1(h^*(t)-h(t))^2}{\int_{0}^1h(t)^2}}\right),
\]
\[
\textrm{fit}_y = 100\left(1-\sqrt{\frac{\int_{0}^1(y^*(t)-y(t))^2}{\int_{0}^1(y(t)-\bar{y})^2}}\right),
\]
\noindent measuring the relative improvement in the quadratic estimation error for the impulse response and the output signal with respect to the baselines $h(t) =0$ and $y(t) = \bar{y} = \int_{0}^1y(\tau)d\tau$, respectively. Figure \ref{FIG03} shows the boxplots of $\textrm{fit}_h$ and $\textrm{fit}_y$ for the 50 experiments, highlighting a significant advantage of $r =2$ at estimating the impulse response, and a slight advantage at fitting the output signal.

\section{Summary}

We have discussed a functional formulation of the LTI system identification problem that allows to handle both discrete and continuous time systems from datasets with arbitrary time sampling, while incorporating several types of structural system properties, such as BIBO stability, relative degree, and smoothness. We have also introduced several examples of kernels, showing that some of them are well-suited to describe stable dynamics while others are not. Finally, we have  outlined the potentialities of applying multiple kernel learning methodologies in the context of LTI dynamical system identification.

\appendix

\section{Proofs}

In this appendix, we provide proofs for all the lemmas presented in the paper.

\textbf{Proof of Lemma \ref{LEM01}:} By the reproducing property, we have
\[
h(t) = \langle K_{t}, h\rangle_{\mathcal{H}},
\]
\noindent If the kernel $K$ satisfies the condition of the Lemma, we have $K_t = 0$ for $t < 0$, so that $h(t)$ equals zero for negative $t$. On the other hand, since $K_{t} \in \mathcal{H}$ for all $t$, condition (\ref{EQUA04}) implies
\[
K_t(\tau) = K(t,\tau) = 0, \quad \forall t < 0, \quad \forall \tau \in  \mathcal{T}.
\]
In view of symmetry, it follows that the kernel must necessarily be in the form defined by the Lemma.
\begin{flushright}
    $\Box$
\end{flushright}

\textbf{Proof of Lemma \ref{LEM02}:}

This is an immediate corollary of Proposition 4.2. in \cite{Carmeli06}.

\begin{flushright}
    $\Box$
\end{flushright}

\textbf{Proof of Lemma \ref{LEM03}:}

If $K$ is integrable, for all $h \in L^{\infty}(\mathcal{T})$, we have
\begin{align*}
& \int_{ \mathcal{T}}\left|\int_{\mathcal{T}}K(t_1,t_2) h(t_1)dt_1 \right| dt_2\\
& \leq \|h\|_{\infty} \int_{ \mathcal{T}} \int_{ \mathcal{T}} |K(t_1,t_2)| dt_1dt_2 < +\infty.
\end{align*}

\begin{flushright}
    $\Box$
\end{flushright}

\textbf{Proof of Lemma \ref{LEM04}:} The proof is similar to that of Lemma \ref{LEM01}. By the reproducing property, we have
\[
h(t) = \langle K_{t}, h\rangle_{\mathcal{H}},
\]
\noindent If the kernel $K$ satisfies the condition of the Lemma, we have $K_t = 0$ for $t < D$, so that $h(t)$ equals zero $t < D$. On the other hand, since $K_{t} \in \mathcal{H}$ for all $t$, condition (\ref{EQUA04}) implies
\[
K_t(\tau) = K(t,\tau) = 0, \quad \forall t < D, \quad \forall \tau \in  \mathcal{T}.
\]
In view of symmetry, it follows that the kernel must necessarily be in the form defined by the Lemma.
\begin{flushright}
    $\Box$
\end{flushright}

\textbf{Proof of Lemma \ref{LEM05}:}

The restriction of the kernel to $\left(0, +\infty\right)^2$ is $k$-times continuously differentiable. Then, by Corollary 4.36 of \cite{Steinwart08}, it follows that the restriction of every function $h \in \mathcal{H}$ to the interval $\left(0, +\infty\right)$ is $k$-times continuously differentiable, and point-wise evaluated derivatives are bounded linear functionals.
\begin{flushright}
    $\Box$
\end{flushright}

\textbf{Proof of Lemma \ref{LEM06}:}

In view of Lemma \ref{LEM05}, point-wise evaluated derivatives at any $t > 0$ are bounded linear functionals. By the reproducing property, we have
\[
\lim_{\tau \rightarrow 0^+}h^{(i)}(\tau) = \lim_{\tau \rightarrow 0^+} \langle K^{(i)}_{\tau}, h\rangle_{\mathcal{H}}.
\]

\noindent If all the impulse responses $h \in \mathcal{H}$ have relative degree greater or equal than $k$, the left hand side is zero for all $h \in \mathcal{H}$ and $i < k$. It follows that
\[
\lim_{\tau \rightarrow 0^+} K^{(i)}_{\tau} = 0, \quad \forall i < k.
\]
\noindent Condition (\ref{EQUA06}) follows from the symmetry of the kernel. Conversely, if condition (\ref{EQUA06}) holds, we immediately obtain
\[
\lim_{\tau \rightarrow 0^+}h^{(i)}(\tau) = 0, \quad \forall i < k, \quad \forall h \in \mathcal{H},
\]
\noindent since the inner product is continuos. It follows that the relative degree of any function $h$ of the space is greater or equal than $k$.
\begin{flushright}
    $\Box$
\end{flushright}

\textbf{Proof of Lemma \ref{LEM07}:}  We have
\begin{align*}
& \int_{\mathbb{R}^2}\left|K(t_1,t_2)\right|dt_1 dt_2 = \\
& = \int_{\mathbb{R}^+}\int_{\mathbb{R}^+ \times \mathbb{R}^+}(t_1t_2)^k\left|G(h_{\omega}(t_1),h_{\omega}(t_2))\right|dt_1dt_2d\mu(\omega)\\
& = \int_{[0, 1]^2}\left(\ln{s_1}\ln{s_2}\right)^k\left|G(s_1,s_2)\right|ds_1ds_2\int_{\mathbb{R}^+}\frac{d\mu(\omega)}{\omega^{2(1+k)}}\\
& \leq \frac{1}{\epsilon^{2(1+k)}} \int_{[0, 1]^2}s_1s_2\left(\ln{s_1}\ln{s_2}\right)^k\frac{\left|G(s_1,s_2)\right|}{s_1s_2}ds_1ds_2\\
& \leq \frac{C}{\epsilon^{2(1+k)}} \int_{[0, 1]^2}s_1s_2\left(\ln{s_1}\ln{s_2}\right)^k ds_1ds_2\\
& \leq \frac{C}{\epsilon^{2(1+k)}} e^{-2k(1-\log{k})} <+\infty.
\end{align*}

The thesis follows by applying Lemma \ref{LEM03}.

\begin{flushright}
    $\Box$
\end{flushright}

\textbf{Proof of Lemma \ref{LEM08}:}
Consider the necessary condition of Lemma \ref{LEM02} and let $h(t) = \cos(\omega t)$ for any $\omega \geq 0$. Let
\begin{align*}
& \phi_{\omega}(t_2) = \int_{\mathbb{R}^+}f(t_1-t_2) \cos(\omega t_1) dt_1 = \\
& =\int_{-t_2}^{+\infty}f(\tau)\cos(\omega (\tau+t_2)) d\tau \\
& = \int_{-t_2}^{+\infty}f(\tau)\left[\cos(\omega\tau) \cos(\omega t_2) - \sin(\omega\tau) \sin(\omega t_2)\right] d\tau.
\end{align*}
\noindent According to Lemma \ref{LEM02}, the kernel $K$ defined by (\ref{EQUA12}) is BIBO stable only if $\phi_{\omega}$ is integrable over $\mathbb{R}^+$. A necessary condition is
\begin{align*}
0 & = \lim_{t_2 \rightarrow +\infty} \phi_{\omega}(t_2) \\
& =  \lim_{t_2 \rightarrow +\infty} \cos(\omega t_2) \int_{-t_2}^{+\infty}f(\tau)\cos(\omega\tau) d\tau,
\end{align*}
\noindent where we have used the fact that $f$ is even, therefore the term containing $\sin(\omega \tau)$ vanishes. Such condition can be satisfied only if
\[
\int_{\mathbb{R}}f(\tau)\cos(\omega\tau)d\tau = 0, \quad \forall \omega \in \mathbb{R}^+,
\]
\noindent which implies $f =0$, and therefore $K =0$.

\begin{flushright}
    $\Box$
\end{flushright}

\end{document}